\documentclass{aa}
\usepackage{graphicx}
\usepackage{txfonts}
\usepackage{natbib}
\bibpunct{(}{)}{;}{a}{}{,}

\newcommand{\eg}    {e.\,g.}
\newcommand{\ie}    {i.\,e.}
\newcommand{\sii}   {\ion{[S}{ii]}}
\newcommand{\nii}   {\ion{[N}{ii]}}
\newcommand{\fe}    {\ion{[Fe}{ii]}}
\newcommand{\oi}    {\ion{[O}{i]}}
\newcommand{\oii}   {\ion{[O}{ii]}}
\newcommand{\ca}    {\ion{[Ca}{ii]}}
\newcommand{\vlsr}  {$V_\mathrm{LSR}$}
\newcommand{\den}  {$n_\mathrm{e}$}
\newcommand{\fri}  {$x_\mathrm{e}$}
\newcommand{\dent}  {$n_\mathrm{H}$}
\newcommand{\kms}   {\mbox{km~s$^{-1}$}}

\begin{document}

\title{The nature of HH~223 from 
long-slit spectroscopy.~\thanks{Based on observations made with the 
2.6~m Nordic Optical Telescope
operated at the Observatorio del Roque 
de los Muchachos of the Instituto de Astrof\' \i sica de Canarias.}
}

\author{R. L\'opez \inst{1}
\and R. Estalella \inst{1}
\and G. G\'omez \inst{2,3}
\and A. Riera \inst{4,1}
\and C. Carrasco-Gonz\'alez\inst{5,6}
}

\offprints{R. L\'opez}

\institute{ Departament d'Astronomia i Meteorologia, Universitat de Barcelona,
Mart\'{\i} i Franqu\'es 1, E-08028 Barcelona, Spain; email: rosario@am.ub.es,
robert.estalella@am.ub.es 
\and Instituto de Astrof\'{\i}sica de Canarias, E-38200,
La Laguna, Tenerife, Spain; email: ggv@iac.es
\and
GTC; GRANTECAN S.A. (CALP), E-38712 Bre\~na Baja, La Palma,
Spain;  email: gabriel.gomez@gtc.iac.es
\and
Dept.\ F\'{\i}sica i Enginyeria Nuclear. EUETI de Barcelona,
Universitat Polit\`ecnica de Catalunya, Comte d'Urgell 187, E-08036 Barcelona,
Spain; email:angels.riera@upc.edu
\and
Instituto Astrof\'{\i}sica Andaluc\'{\i}a, CSIC, Camino
Bajo de Hu\'etor 50, E-18008 Granada, Spain; email: charly@iaa.es
\and 
Centro de Radiastronom\' \i a y Astrof\' \i sica UNAM, Apdo Postal 3-72
(Xangari), 58089 Morelia, Michoac\'an, M\'exico.
}

\date{\today}

\titlerunning{Long-slit spectroscopy HH~223 }

\abstract{HH~223 is a knotty, wiggling nebular emission of $\sim 30\arcsec$ length
found in the L723 star-forming region. It lies projected onto the largest
blueshifted lobe of the cuadrupolar CO outflow powered by a low-mass YSO system
embedded in the core of L723.}
{We analysed the physical conditions and kinematics along HH~223 with the 
aim of disentangling whether the emission arises from shock-excited, supersonic
gas characteristic of a stellar jet, or is only tracing the wall cavity
excavated by the CO outflow.}
{We performed long-slit optical spectroscopy along HH~223,  
crossing all the bright knots (A to E) and 
part of the low-brightness emission nebula (F filament). One spectrum of each
knot, suitable to characterize the nature of its emission, was obtained. The physical conditions and the radial velocity of the HH~223 emission
along the slits were also sampled at smaller scale ($0\farcs6$) than the 
knot sizes.}
{The spectra of all the HH~223 knots appear as those of the
intermediate/high excitation Herbig-Haro objects. The emission is supersonic,
with blueshifted peak  velocities ranging from $-60$ to $-130$ \kms. Reliable
variations  in the kinematics and physical
conditions at smaller scale that the knot sizes are also found.}
{The properties of the HH~223 emission derived from the spectroscopy confirm the
HH nature of the object, the supersonic optical outflow  most probably also being
powered by the YSOs embedded in the L723 core.}


\maketitle

\section{Introduction}

\object{Lynds 723} (L723) is an isolate dark cloud located at a distance of 300$\pm$150
pc  (Goldsmith et al.~\citealp{gol84}) that shows evidence of low-mass star
formation, where one of the few known cases of a
quadrupolar CO outflow (two separate pairs of red-blue lobes; Lee et
al.\ \citealp{lee02}, and references therein) has been reported.
The 3.6~cm radio  continuum source \object{VLA~2} (Anglada et
al.\ \citealp{ang96}), towards the centre of the CO outlfow, seems to harbour
the  exciting outflow source. 
VLA~2 is embedded in high-density gas traced by NH$_3$,
showing evidence of gas heating and line broadening (Girart et al.\ 
\citealp{gir97}). The dense envelope of the source has also been observed at
submillimeter wavelengths by Shirley et al.\ \cite{shi02}, and Estalella et
al.\ \cite{est03}.
Recent works by Carrasco-Gonz\'alez et al.\ \cite{car08} and Girart et al.\
\cite{gir08},
show that VLA~2 is a multiple system.
Carrasco-Gonz\'alez et al.\ \cite{car08} detect at least four young stellar
objects (YSOs) and propose that the morphology of the CO outflow is
actually the result of the superposition of three independent
pairs of CO lobes.
They also propose
that one of the YSOs (\object{VLA 2A}) is exciting the largest pair of CO
lobes and the system of emission-line nebulosities, reminiscent of Herbig-Haro
(HH) objects, first 
reported by Vrba et al.\ \cite{vrb86}.

\begin{figure*}
\centering
\resizebox{0.8\hsize}{!}{\includegraphics{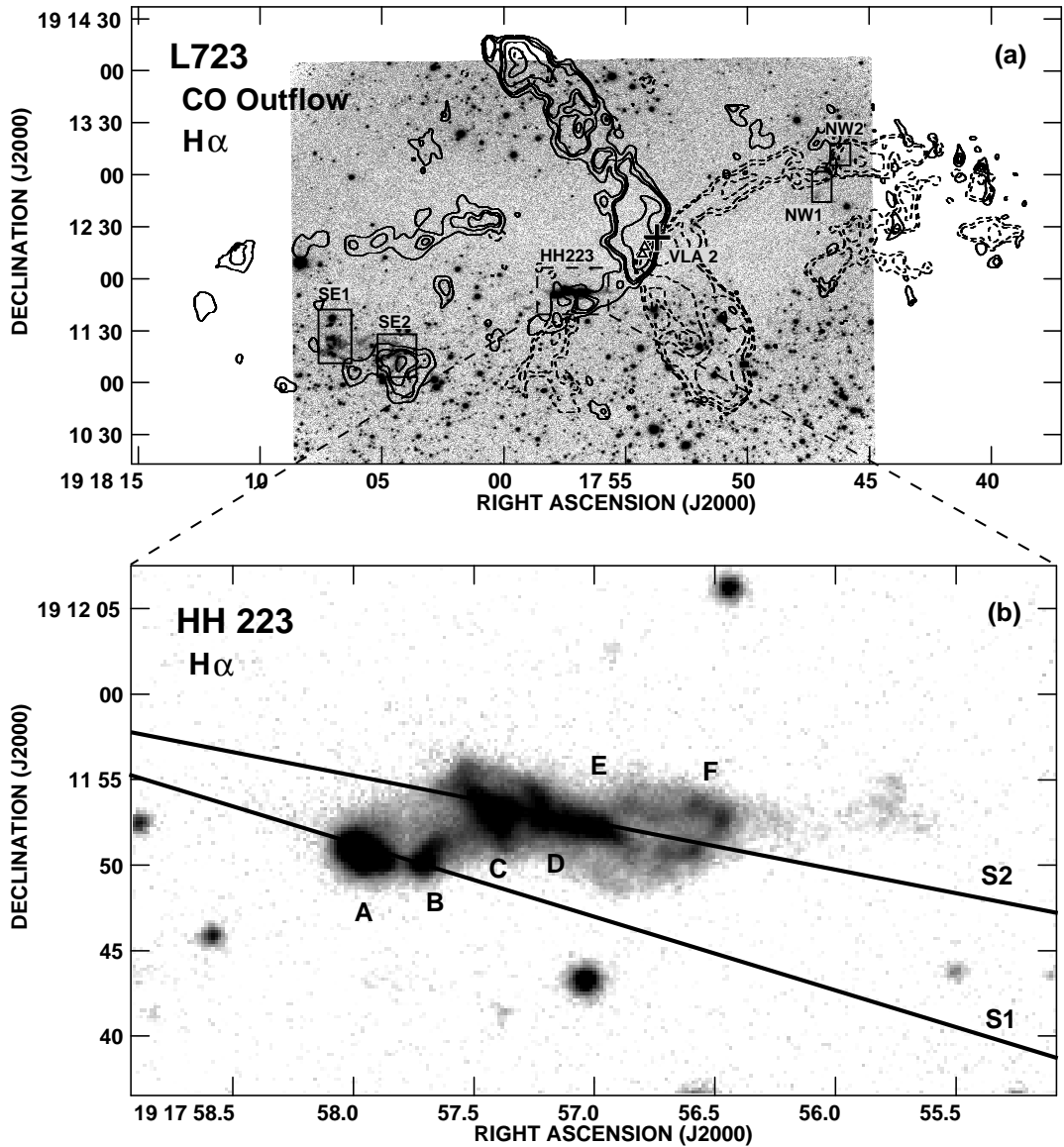}}
\caption{(a) H$\alpha$ image of the L723 field. HH~223 and the other H$\alpha$
nebulosities given in L\'opez et al.\ \cite{lop06} have been labeled and
enclosed within boxes. The triangles mark the positions of the H$_2$ knots K1
and K2 of Palacios \& Eiroa \cite{pal99}. Contours of the high-velocity 
CO (\mbox{$J$=1$\rightarrow$0}) emission from Lee et al. \citealp{lee02}
have been superposed (the dashed contours correspond to redshifted emission and
the solid contours to blueshifted emission). The VLA2 source (Anglada et al.\ 
\citealp{ang96}) is marked. (b) Close-up of the image showing
HH~223. The slit positions for the spectra
acquisition (S1, S2) are indicated. Knots are labeled according to the nomenclature of 
L\'opez et al.\ \cite{lop06}.}
\label{image} 
\end{figure*}

In a previous work (L\'opez et al.\ \citealp{lop06}), we presented 
deep narrow-band H$\alpha$ and \sii\
images of L723, which confirmed the emission-line nature of the nebulosities. 
 The H$\alpha$ image, with the CO outflow contours  from 
Lee et al.\ \cite{lee02} superposed, is shown in panel (a) 
of Fig.\ \ref{image}.
The higher angular resolution of the images allowed us 
to resolve
the detailed structure of \object{HH~223}, the brightest ''linear emission feature''
reported by Vrba et al.\ \cite{vrb86}. HH~223   appeared
as a  set of knots embedded in a fainter nebula, following a wiggling pattern,
reminiscent of a stellar jet, that is probably  part of a parsec-scale HH 
outflow  as suggested by CO observations of Lee et al.\ \cite{lee02},
optical (L\'opez et al.\ \citealp{lop06}) and near-infrared
(Palacios \& Eiroa \citealp{pal99}; L\'opez et
al.\ \citealp{lop09}) imaging of the shocked emission.
However, there were no
kinematic data  to  establish the nature of the line emission,
\ie\ whether the emission is supersonic (as in typical stellar jets) or, in contrast, is
nearly stationary as would be expected  for a wall cavity. With the aim of
exploring the kinematics and physical conditions in HH~223, we performed
long-slit spectroscopy of the object covering all  its bright knots (A to E) and
a part of the low-brightness nebula (F filament). The results  are 
presented in this paper.

\section{Observations and Data Reduction}

\begin{figure*} 
\centering
\rotatebox{-90}{\includegraphics[width=0.7\hsize,clip]{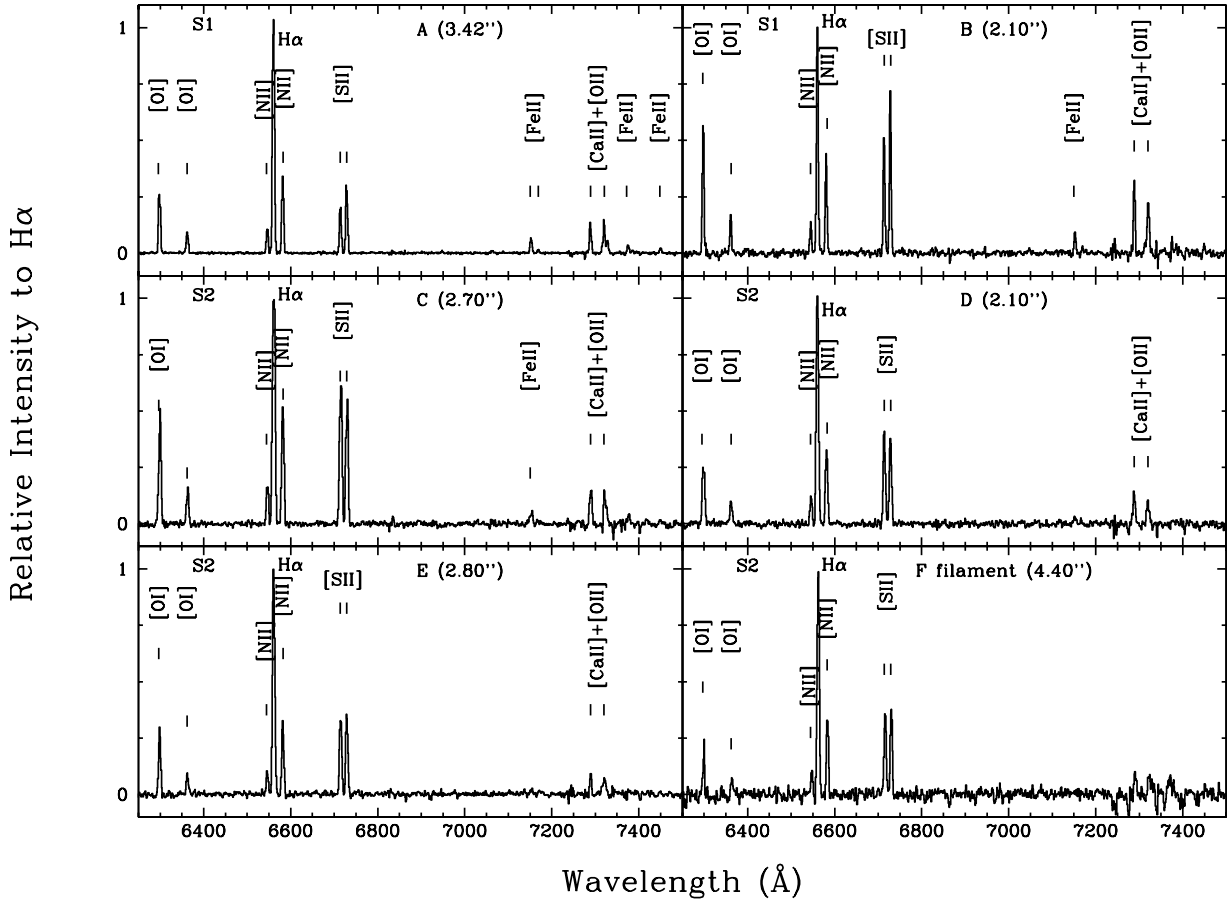}}
\caption{Spectra of the HH~223 knots from east (A, top-left panel)
to west (F filament, bottom-right panel)
obtained by averaging the signal inside the slit crossing the knot. In all the
spectra, the intensity has been normalized to the H$\alpha$  peak intensity. The
emission lines detected  in each  spectrum have been labeled in the panels. The
slit position (and the size of the  region averaged along the slit to get the 
spectrum) are indicated in each panel.}
\label{espint}  
\end{figure*}

Spectra of HH~223 were obtained  with the 2.6-m Nordic Optical Telescope (NOT)
of the Observatorio del Roque de los Muchachos (ORM, La Palma, Spain) on 13 July
2007. The  ALFOSC (Andaluc\' ia  Faint Object Spectrograph and Camera), with an
image scale of $0\farcs188$~pixel$^{-1}$, was used in its spectroscopic mode.
The 600R grism \# 8, covering the wavelength range  5825-8350 \AA\ with a
dispersion of 1.3 \AA\ pix$^{-1}$ (spectral resolution $\simeq$~60~\kms)  was
used. In order to cover all the  HH~223 bright knots, the spectra were acquired
through a long-slit of 6$\farcm$5 length  and  1\arcsec\ width, positioned  at two
position angles (see panel (b) of Fig.\ \ref{image}): at PA 73$^\circ$, along 
knots A and B (S1); and at  PA
79$^\circ$, along knots C to F (S2).  Individual
exposures of 1800~s each were made to get total exposure times of 5400~s and
7200~s for the S1 and S2 slit positions, respectively. 
Special care was taken in
achieving accuracy to  point the slit on the  knots. The setup only allows
horizontal or vertical slit positions.  Thus, a shorter exposure (200~s) of the
field oriented at the desired PA was previously acquired through a narrow-band
H$\alpha$ filter.  Then, an image of the slit  on the detector was made and the
procedure was  iterated twice to achieve an accurate slit position  before
acquiring the
spectrum.  Seeing was $0\farcs6$ for  most of the run. The
data were processed with the standard tasks for long-slit spectroscopy  of the
IRAF\footnote{IRAF is distributed by the National Optical Astronomy
Observatories, which are operated by the Association of Universities for
Research in Astronomy, Inc., under cooperative agreement with the National
Science Foundation.} package,  which include bias substraction, flat-fielding
correction, wavelength calibration using spectra from a Ne lamp acquired at
several elevations for each  slit position,  and sky substraction. The achieved
accuracy for the wavelength calibration, checked from the sky lines, was better
than $\sim$ 0.2 \AA\ ($\sim$ 10 km~s$^{-1}$).  The accuracy in the determination
of the position of the line centroids  is $\sim$ 0.2 \AA\ ($\sim$
10~km~s$^{-1}$) for the strong  emission lines observed.  The final spectrum
through  each slit position was corrected for cosmic ray events by median
filtering all the exposures, after  wavelenght calibration, obtained at the
given slit position. The spectra were not flux calibrated.

\section{Results and Discussion}

\subsection{The spectra of the HH~223 knots: Physical conditions and kinematics}

We obtained the spectrum for each HH~223 knot defined in L\'opez et al.  \cite{lop06}
by averaging the emission within the slit window
crossing the knot.
The six spectra obtained in such a way are displayed in Fig. \ref{espint}.  
As can be seen from
the figure, the characteristic HH emission lines from \oi\ $\lambda$~6300, 6364
\AA , \nii\ $\lambda$~6548, 6583 \AA , H$\alpha$ and \sii\ $\lambda$~6716, 6731 \AA\
were detected in all the knot spectra. In addition, other emission lines
redwards of 7000 \AA\ were detected in several knots:  the \fe\ 14F 
multiplet lines
($\lambda$~7155, 7173, 7388, 7453 \AA ) were detected in knot A; the brightest
\fe\ $\lambda$~7155 \AA\  component of 14F multiplet was also detected in knots
B and C, and the  \oii\ $\lambda$~7321, 7331 \AA\ and \ca\  $\lambda$~7291, 7324
\AA\ lines  (the two doublets being blended) were detected in knots A to E.

\begin{table*}
\caption[ ]{\vlsr\ line centroids$^{1}$ of the HH~223 knots}
\label{tvknot}
\begin{tabular}{cccccc}  
\hline\hline
 Knot& \oi\ 6300& H$\alpha$& \nii\ 6583& \sii\ 6731& Average\\
\hline 
A  &\phantom{1}$-80\pm 20$&$-108\pm 15$&\phantom{1}$-83\pm 15$&$-108\pm 15$&\phantom{1}$-95\pm33$\\
B  &$-115\pm 15$&$-122\pm 15$&$-131\pm 15$&$-128\pm 15$&$-124\pm 30$\\
C  &\phantom{1}$-23\pm 15$&\phantom{1}$-76\pm 15$&\phantom{1}$-67\pm 15$&\phantom{1}$-57\pm 15$&\phantom{1}$-56\pm 30$ \\
D  &\phantom{1}$-80\pm 15$&$-117\pm 15$&\phantom{1}$-92\pm 15$&$-106\pm 15$&\phantom{1}$-99\pm 30$ \\
E  &\phantom{1}$-63\pm 15$&\phantom{1}$-99\pm 15$&\phantom{1}$-79\pm 20$&\phantom{1}$-79\pm 25$&\phantom{1}$-80\pm 33$ \\
F  &\phantom{1}$-24\pm 15$&\phantom{1}$-26\pm 15$&\phantom{11}$-1\pm 15$&\phantom{1}$-17\pm 25$&\phantom{1}$-17\pm 35$ \\
\hline
\end{tabular}
\begin{list}{}{}
\item[] Notes:\\
(1) Derived from the spectra integrated within the slit 
aperture sizes indicated  in Fig.\ \ref{espint}.\\
\end{list}
\end{table*}

\subsubsection{Density and excitation conditions of the knots}

From the spectra obtained as mentioned above, 
we  explored the electron density (\den), from the \sii\ 6716/6731 line 
ratio, and
the excitation conditions, from the \nii/H$\alpha$ and \sii/H$\alpha$ line
ratios.  The fluxes were uncorrected for reddening. However,
these line ratios are insensitive to differential extinction because of
the wavelength proximity of the lines.

\begin{figure}
\centering
\includegraphics[width=\hsize,clip]{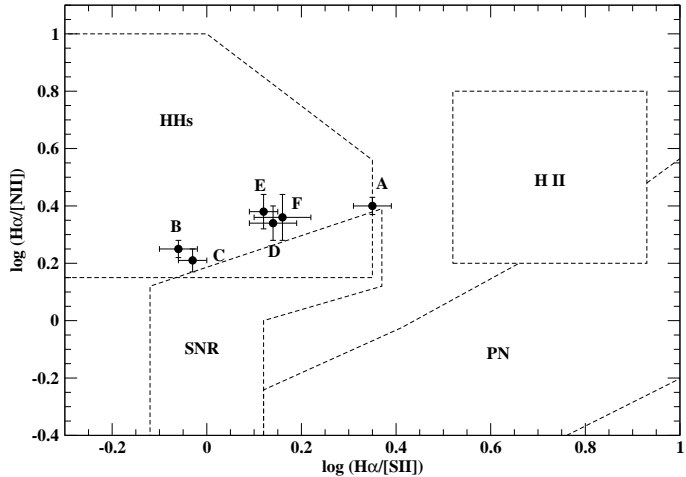}
\caption{Position of the HH~223 knots in a diagnostic diagram 
of H$\alpha$/\sii\ vs.\ H$\alpha$/\nii\ (line ratios, obtained from the
knot spectra of Fig.\ \ref{espint}). The regions of the Planetary
Nebulae (PN), Supernova Remnants (SNR), H~II regions (H~II) and Herbig-Haro
objects (HHs), adopted from the figures by  Cant\'o \cite{can81}, are outlined.}
\label{diag1} 
\end{figure}

\begin{figure}
\centering
\includegraphics[width=\hsize,clip]{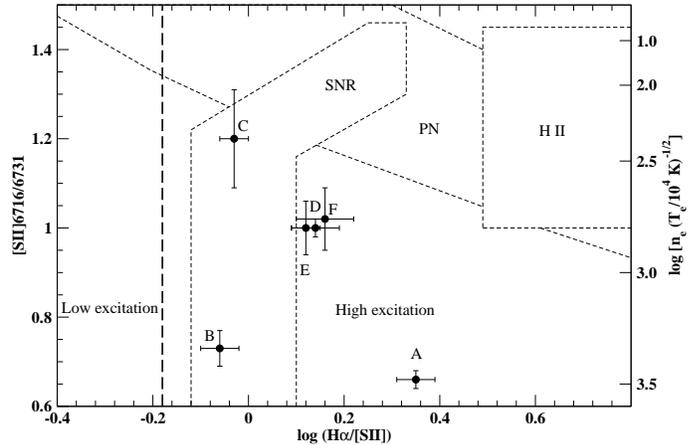}
\caption{Position of the HH~223 knots in the excitation vs.\ density 
diagnostic diagram. 
The long-dashed line gives the separation between high/intermediate and low
excitation HH spectra (see Raga et al. \citealp{rag96}). 
The regions of the Planetary
Nebulae (PN), Supernova Remnants (SNR) and H~II regions (H~II), adopted 
from the figures of Meaburn \& White \cite{mea82}, are outlined.}
\label{diag2} 
\end{figure}

Figures \ref{diag1} and \ref{diag2}  show the location of the HH~223 knots in
current diagnostic diagrams from line ratios of nebular  emission lines (see
\eg\ Cant\'o \citealp{can81}; Raga et al. \citealp{rag96}). As can be seen
from these figures,
the line ratios of all the HH~223 knots lie in 
the diagrams region where are located the HH objects having an intermediate- or
high-excitation spectrum. The
diagrams also show that there are appreciable differences in both, excitation and density,
among  the knots. 
Knot A  has the highest excitation  spectrum. Concerning the rest of the
HH~223 knots,  there is some trend of increasing excitation going towards 
the west. 
Values for \den\ of the knots  were derived  from the \sii\ 6716/6731 ratio,
using the {\small TEMDEN} task of the {\small IRAF/STSDAS} package, and
assuming $T_\mathrm{e}=10^4\ \mathrm{K}$.  The results are shown in 
Fig.\ \ref{diag2}.
The electron
density is significantly higher in the brighter, southern knots A and B  than in
the rest of the HH~223 knots,  the difference between the highest 
(in knot A) and the lowest 
(in knot C) \den\ values being about one order of
magnitude. Knots D, E and F have similar  \den\ values. Note, however, that knots B and C 
have spectra of  similar excitation,
in spite of  their derived \den\ being significantly different.

\begin{figure}
\centering
\resizebox{0.72\hsize}{!}{\includegraphics{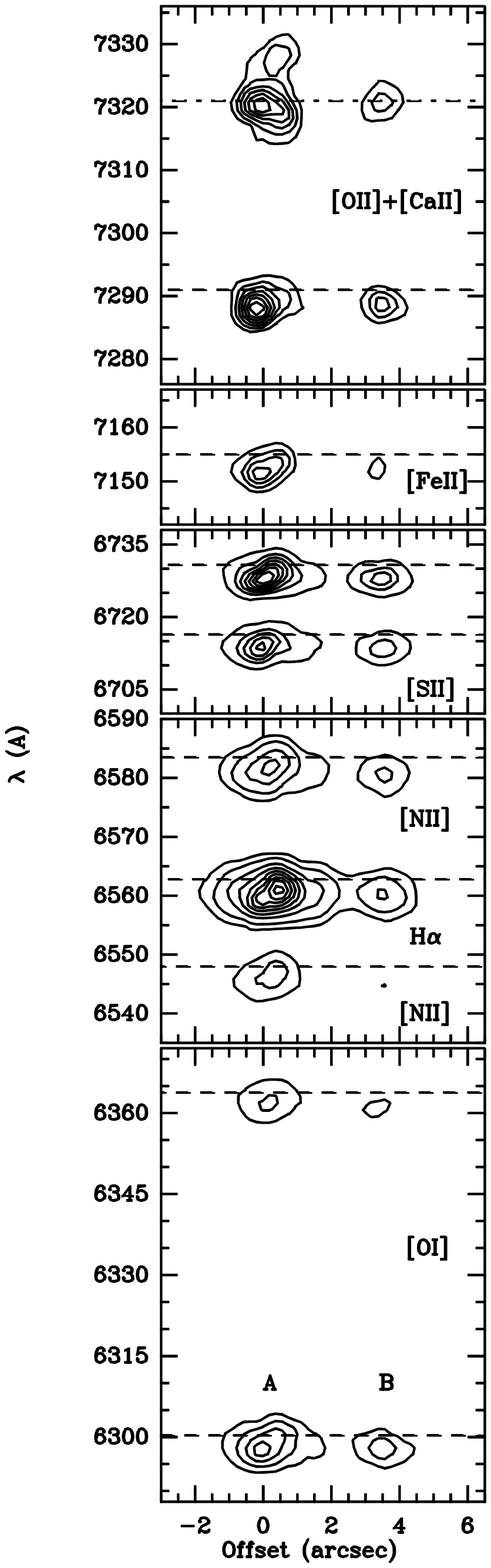}}
\caption{Wavelength-position diagram of the HH~223 emission 
along slit position S1 (PA 73$^\circ$) for the lines
indicated in each panel. The rest 
wavelength that corresponds to  \vlsr = 0~\kms\ is
marked with a dashed line. On the \oii\ + \ca\ panel, the  rest
wavelengths marked correspond to the \ca\ $\lambda$ 7291 \AA\ (dashed)
 and to the \oii\ $\lambda$ 7321
\AA\ (dotted-dashed) component of their respective doublets. 
The reference offset
position corresponds to the \sii\ knot A peak intensity. 
The knots identification is
labeled in the \oi\ panel.} 
\label{plab}
\end{figure}

\begin{figure}
\centering
\rotatebox{-90}{\includegraphics[width=1.5\hsize,clip]{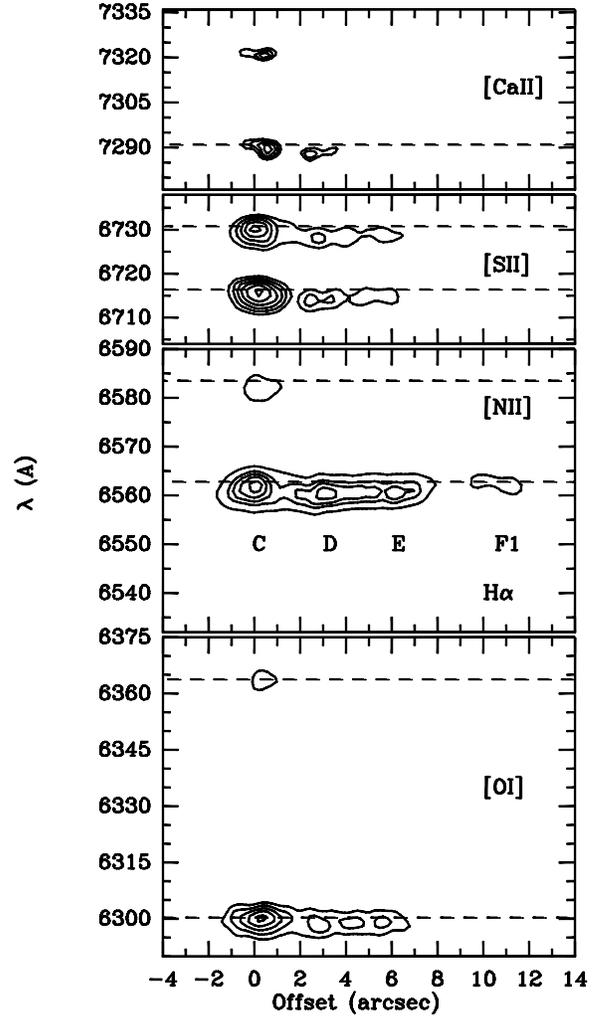}}
\caption{Same as Fig.\ \ref{plab}, but for the slit position S2 (PA 79$^\circ$).
The reference offset
position corresponds to the \sii\ knot C peak intensity. 
The knots identification is labeled in the H$\alpha$ panel.}
\label{plcf}
\end{figure}

\begin{figure}
\centering
\resizebox{0.85\hsize}{!}{\includegraphics{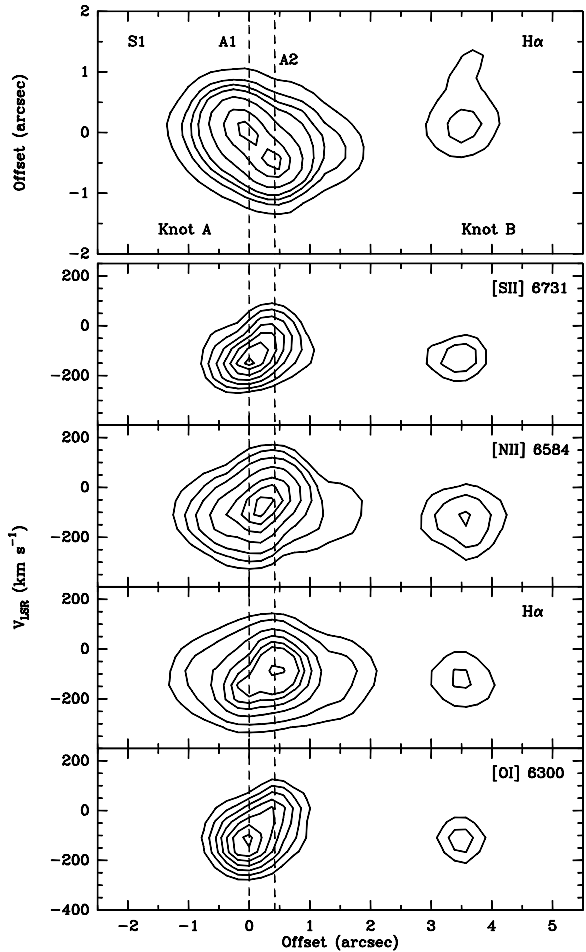}}
\caption{Upper panel: Contour plot of the H$\alpha$ emission in knots A and B
obtained from a  narrow-band image of the L~723 field.
Lower panel: 
PV maps for different emission lines, labeled in each panel.
Contours are in counts (image and spectra are not flux calibrated)
and the contour spacing is 10 per cent of the knot A peak intensity.
All the panels are oriented at a PA=73$^\circ$ (corresponding to the S1 slit
position, crossing the HH~223 knots A 
and B in the east-west direction). The reference offset
position is as in Fig.\ \ref{plab}. The dashed lines mark the peak 
positions in the PV maps of the two substructures, A1 and A2, of knot A.}
\label{pvab}
\end{figure}

\subsubsection{Radial Velocities of the knots}

The  radial velocities\footnote{All the velocities in the paper are referred to
the local standard of rest (LSR) frame. A   \vlsr=+10.9~\kms\ for the parent cloud
has been taken from Torrelles et al. \citealp{tor86}} of the HH~223 knots were obtained  
from the line
centroids of  Gaussian fits to the  emission lines of the  spectra of 
Fig.\ \ref{espint}. Thus, the derived velocities correspond to the  knot
emission integrated within the slit aperture  and do not account for differences in
velocity at spatial scales smaller than the knot size. In Table \ref{tvknot} 
we list the velocities obtained from several emission lines and the average
velocity. The velocities derived for the knots appear blueshifted relative  to
the ambient gas, the gas intersected by the slit moving supersonically towards
the observer. Thus, the HH~223 knot emission properties are those expected from an
optical  outflow. Note that the velocity derived for HH~223-F is significant
lower.  However, the spectrum of HH~223-F mostly arises from gas of the
low-brightness nebula in which the knots are engulfed (F filament) instead of
from a truly, compact knot. 

More detailed  maps  are shown 
in  Figs.\ \ref{plab} and \ref{plcf}. For knot A, 
the  position-wavelength (PV) maps of Fig.\ \ref{plab}  show  a velocity gradient within
this knot, with higher blueshifted  velocities  towards the east, the trend
being found in all the  lines mapped.  In fact, the  PV map of  the H$\alpha$
line suggests that there are  two velocity peaks, at \vlsr $\simeq$ 
$-145$ and $-85$  \kms,  with a position offset $\simeq$~ 0$\farcs$5.  
Interestingly, the
narrow-band H$\alpha$ image  also shows two
intensity maxima  within
knot A (labeled A1 and A2 in Fig.\ \ref{pvab}, upper panel), the positions of 
these intensity maxima being in good concordance with
the positions of the two velocity components of the PV maps.  
This suggests that the emission labeled as knot A, extending over $\sim$
3$\farcs$5  along the slit position S1, would be in fact a more complex structure
composed of two emitting clumps, partially resolved by our data,  each  clump
moving at a different blueshifted velocity. 
This is better seen in 
Fig.\ \ref{pvab}, where we show the PV maps for different lines
(lower panels) and  the intensity contours
obtained from the H$\alpha$ image, oriented according to the slit position S1.
Note in addition that the peak
velocity position in the H$\alpha$ line (\ie\  around the A2 substructure) is
displaced from  the peak velocity in the \sii\ $\lambda$ 6731 \AA\  line (\ie\
around the A1 substructure). Spatial displacements between the H$\alpha$ and  \sii\
peak  positions are also found in other stellar jets (\eg\ 
HH~110, Riera et al.\ \citealp{rie03}; HH~119, G{\aa}lfalk \& Olofsson
\citealp{gal07}), in which the supersonic jet outflow propagates in an
inhomogeneous, quite dense ambient environment.

\subsection{Small-scale properties of the HH~223 emission}

\begin{figure*}  
\centering
\rotatebox{-90}{\includegraphics[width=0.7\hsize,clip]{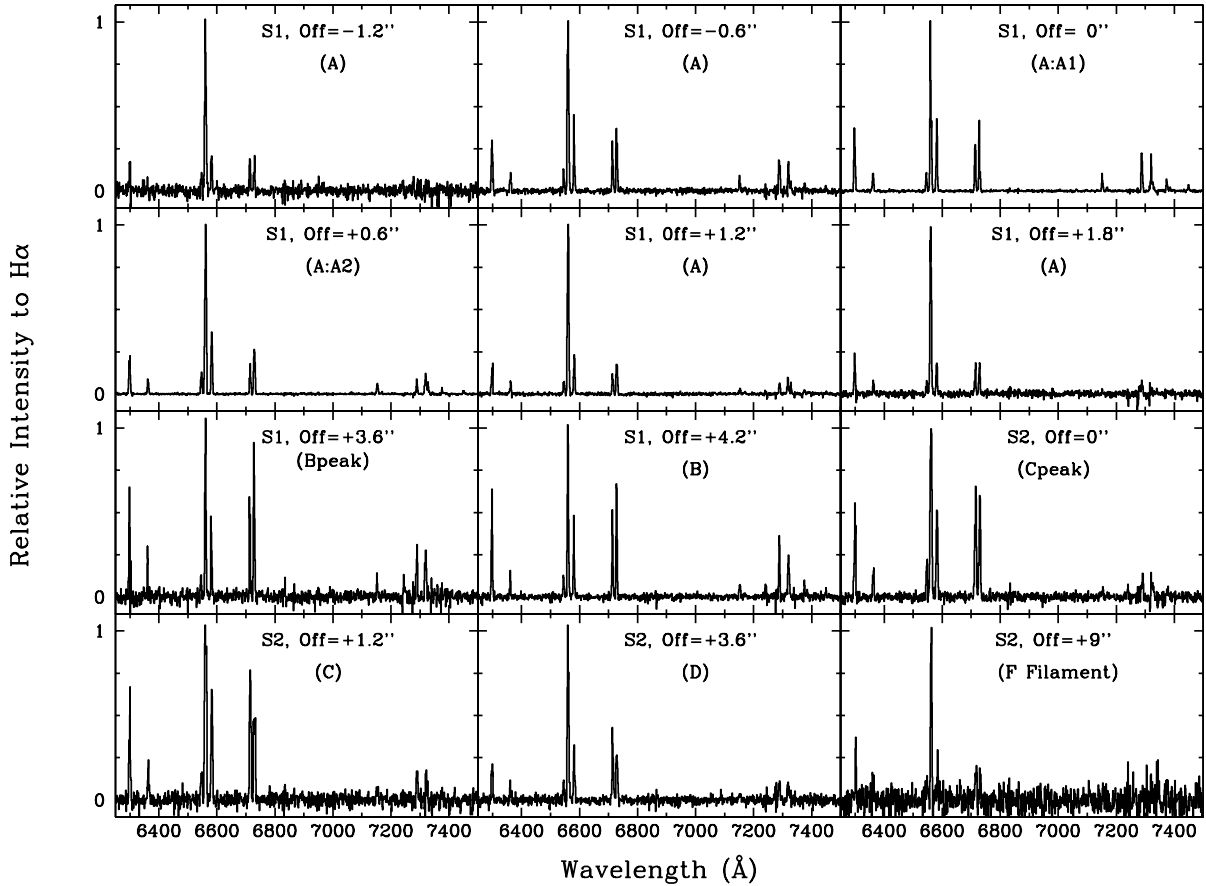}} 
\caption{
Spectra at selected positions along the slits, from east (top left) to west 
(bottom right), belonging to the knots labeleld in each panel. Each spectrum was
obtained by binning the signal within a window of  three
pixels ($\sim$ 0$\farcs$6) centered at the position labeled in the
panel. The  reference postition  (Off=$0''$) for the slit  S1 
corresponds to the 
\sii\ peak intensity position of knot A, which coincides with the substructure A1.
The  reference postition  (Off=$0''$) for the  slit  S2 corresponds to the \sii\ peak
intensity position of knot C. In all the spectra, the intensity has been
normalized to the H$\alpha$ peak intensity.}
\label{espab}  
\end{figure*}

Our data, having better spatial resolution than the knot sizes
(ranging from $\sim$ 2\arcsec\ to 3$\farcs$5), suggest that there
are changes in both kinematics and physical conditions at a smaller 
spatial scale. With the aim of looking for such variations, we
extracted a set of spectra  by  co-adding the signal of each three adjacent
pixels  (\ie\ by doing a spatial binning of $\sim$ 0$\farcs$6, of the order of
the seeing) to cover without gaps  all the HH~223 emission intersected by the  
slits. The spectra, still keeping  a very good
signal-to-noise ratio (several of them are displayed in  Fig.\
\ref{espab}) are then useful for  sampling the  gas conditions  
with a step of $\sim$ 0$\farcs$6. 

The spectra displayed in the two upper rows of  Fig.\ \ref{espab} correspond to
the emission at  adjacent positions through knot A (from east to
west).  A quick look to the panels shows apreciable changes in the relative
intensity of the emission lines, suggestive of changes in excitation and
electron density even by only  moving  $\sim$ 0$\farcs$6 along the slit.  Fig.\
\ref{espab}  also displays some of the spectra obtained at several selected
positions along slits S1 (within knot B) and S2,
for positions belonging to knots C, D and the  F filament. It should  also be
noted that  the \sii\ $\lambda$ 6716 \AA\ line appears
brighter  than the  $\lambda$ 6731 \AA\ line in the spectra extracted along S2, 
while the opposite occurs  along S1. Given that the 
\sii\ $\lambda$ 6716/6731 line ratio is a tracer of the density of the ionized
gas, this finding is indicative of a general trend,  both from the knots and
from the  interknot low-brightness regions, of the gas being less dense 
towards the northwest, \ie\ towards  the optical obscured
region. 

Let us now examine the small-scale changes in the physical conditions 
using the line ratios derived from these spectra. 

\subsubsection{Small-scale excitation and density along HH~223}

\begin{figure}
\centering
\resizebox{0.95\hsize}{!}{\includegraphics{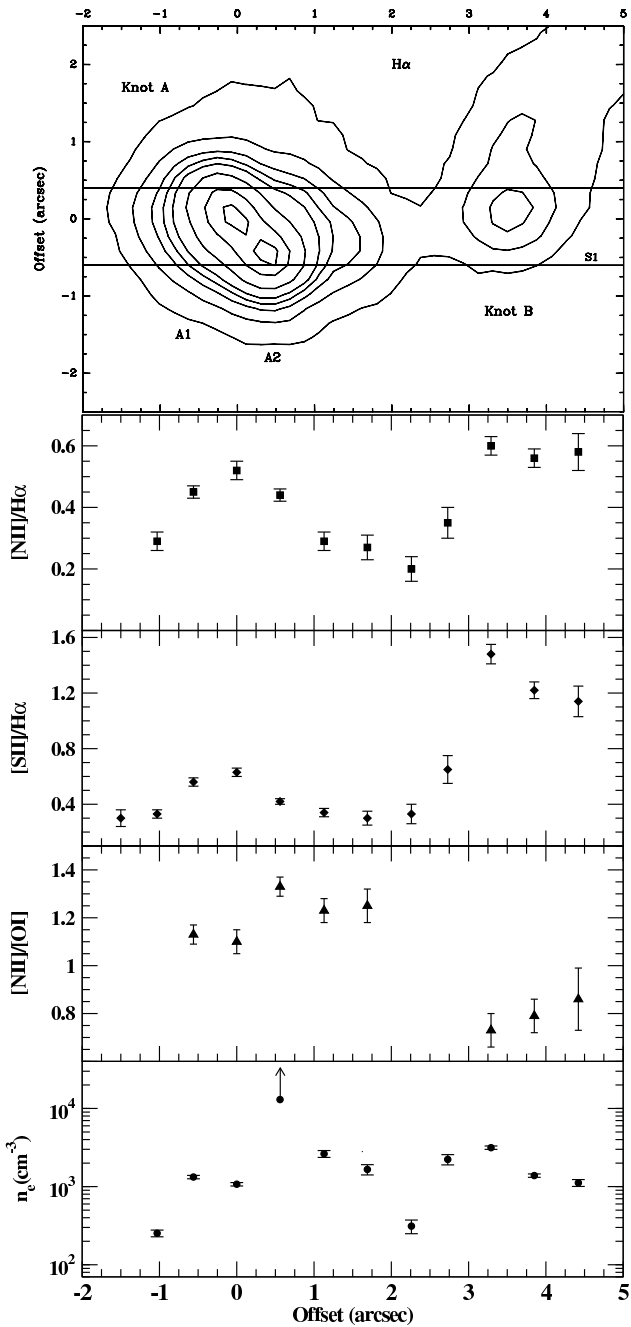}}
\caption{
Physical conditions along  knots A and B (slit position S1).
Upper panel: Contour plot of the H$\alpha$ emission, as in Fig.\ \ref{pvab}. 
The slit position has been marked by the horizontal lines.
Lower panel, from top to bottom: excitation (\nii/H$\alpha$, \sii/H$\alpha$),
ionization  (\nii/\oi), and \den\ (in logarithmic scale, obtained from
the \sii\ $\lambda$ 6716/6731 ratio), as a function of the 
position along the slit. The reference position 
for the offsets is as in Fig.\ \ref{plab}.}
\label{diagab} 
\end{figure}

\begin{figure}
\centering
\resizebox{\hsize}{!}{\includegraphics{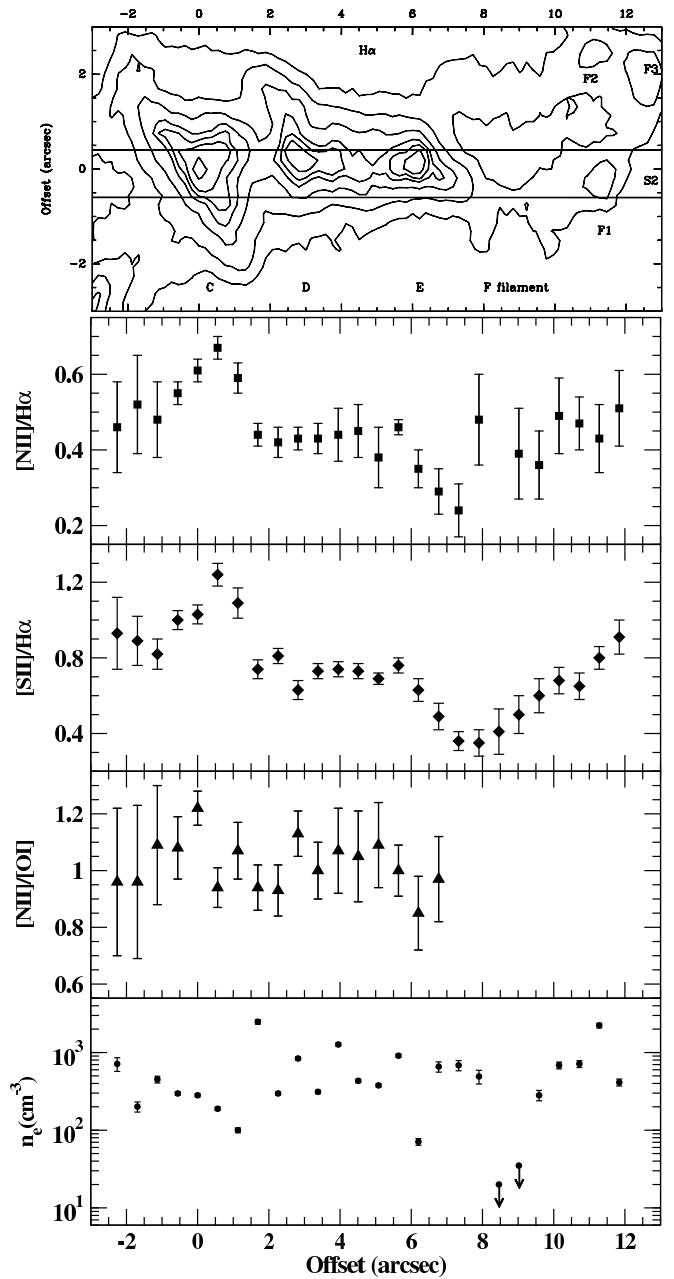}}
\caption{Same as Fig.\ \ref{diagab} but for the slit position S2,
intersecting emission from knots C to F. The reference position for 
the offsets is as in Fig.\ \ref{plcf}.}
\label{diagcf}
\end{figure}

Figures\ \ref{diagab} and \ref{diagcf} (lower panels) display the spatial
behaviour  along the slit for several line ratios, tracing the excitation and
ionizaton conditions,  and for the electron density.
To help with the knot
identification, the H$\alpha$ image is displayed in
the upper panel of these figures. 

As  can be seen from Fig.\ \ref{diagab}, all the line ratios differ from knot to
knot, but also vary at a smaller spatial scale, within the knot.  The gas
excitation is traced by the \nii/H$\alpha$  and \sii/H$\alpha$ line ratios.  All
the  values derived  along knots A and B correspond to the
intermediate/high excitation HHs  (Raga et al. \citealp{rag96}). However, the
excitation is clearly different in both knots: the \sii/H$\alpha$ line ratios
along knot A are  $\leq$ 0.6, while they are $\geq$ 0.9 along knot B. Note 
in addition that
the lowest \sii/H$\alpha$ line ratio value (thus the highest excitation) 
is found inside knot A, at positions offset $\geq\pm 1''$ from the 
two knot A
peaks. In contrast, the \sii/H$\alpha$ line ratio values of knot B are the
highest derived for all the sampled emission, thus indicating that  the lowest
excitation in the emission is found around knot B. These results are fully 
consistent with those derived by L\'opez et al.\ \cite{lop06} using the
\sii/H$\alpha$ ratio map obtained from the narrow-band CCD images.

The  behaviour of the electron density through
knot A and B, derived from
the \sii\ $\lambda$ 6716/6731 line ratio (see Section 3.1.1), 
is also shown in Fig.\ \ref{diagab}. An electron
density 
\den\  $\simeq$ 200-300 cm$^{-3}$ is found at positions away from the knot peaks,
coinciding with the regions where the slit mainly intersects  gas coming from
the low-brightness emission. In contrast,  values in the
range 1200 $\leq$ \den\ $\leq$ 3000 cm$^{-3}$ are found  through knot B and 1500
$\leq$ \den\ $\leq$ 2500  cm$^{-3}$ through knot A, except for the position
around the A2 substructure, where  the derived \sii\ 6716/6731  line ratio is
close to the limit for which this ratio is not a reliable
indicator of the density. A more accurate estimation of \den\ in this knot
should be  derived from the \fe\ 7155/8617 line ratio. However, although the
\fe\ $\lambda$~7155 \AA\ line is detected in  knot A,  unfortunately
the \fe\ 
$\lambda$~8617 \AA\ line is not included within the spectral range used. It should be remarked that such a  high \den\  is most
probably real (\ie\ is not an artifact: first, because a similar
low ratio has been found in each of the three adjacent pixels  binned to get
the  A2 spectrum; and second, because the same behaviour at this knot position 
was obtained  from another exploratory spectrum acquired one year before). Thus, this result 
gives also support to the idea that
the knot A emission is split into two different
substructures that are not well resolved by our data. 

A similar analysis, but for the slit position S2, 
intersecting emission from knots C to F,
is shown in Fig.\ \ref{diagcf}. Regarding the gas excitation, the 
\nii/H$\alpha$ and
\sii/ H$\alpha$ line ratios suggest that three different regions should be
defined along the 
slit: the first one, extending up to $\sim\pm$ $1\farcs5$  
from the knot
C peak, having the  highest \sii/H$\alpha$ ratios ($\geq$ 1)  (the lowest
gas excitation found along S2, being however  higher  than  the  excitation
infered in  knot B); the second one extending up to 7\arcsec\ west beyond
the knot C intensity peak and enclosing  knots D and E, where  the excitation
increases, although it is lower than in knot A; and
finally, the western region beyond the end of knot E, which includes the F 
filament
and the F1 faint knot, where the gas  excitation decreases again.
The spatial behaviour of the electron
density (derived from the \sii $\lambda$ 6716/6731 ratio) along knots C to F
is displayed in the bottom panel of the figure. The
trend found is sugestive of
density inhomogeneities in the  emitting gas, the sizes of the clumps remaining
unresolved with our spatial resolution. As in the case of the knots A and B 
emission, we found two densitiy gas components: a lower-density emitting gas,
with 100 $\leq$ \den $\leq$ 300 cm$^{-3}$ and a higher-density component, with 
700 $\leq$ \den $\leq$ 2500 cm$^{-3}$.  A more careful inspection of the  \den\
spatial distribution also shows that  there are several density enhancements 
within each knot, possibly indicating that the knots could be split in
substructures by  imaging the emission with  higher angular resolution.
 The
\nii/\oi\ line ratio only depends weakly on the temperature, being a good
indicator of the level of global ionization in the flow 
(see \eg\ Bacciotti \&
Eisl\"offel \citealp{bac99}). Its values as a function of the slit position are
plotted in Figs. \ref{diagab} and \ref{diagcf} to qualitatively show the
variation of the degree of ionization in the outflow. As can be seen from
these figures, this ratio is appreciably higher ($\geq$ 1) along knot A than
along knot B, (where the ratio is $<$ 1, the lowest value found for all the
knots). Thus, knot A  appears as the most excited and highly
ionized region in HH~223, while the closest region, knot B,  
is the lowest ionized and excitated one. From the Hartigan, Morse \& Raymond \cite{har94} models, the \nii/\oi\
line ratios allowed us to constraint the ionization fraction (\fri ) to 25--30\%
along knot A, 15\% along knot B and 20--25\% along knots C to E, all these \fri\ values
being roughly independent of the range of preshock magnetic field strengths
expected within molecular clouds. The \nii/\oi\  ratios also constraint the
shock velocities to 60--70 \kms. From the estimated \fri\ values and the derived
\den , we  also
estimated the density (\dent = \den/\fri) for the knots. We found that knots A
and B are significantly denser ( \dent $\ge$ 10$^4$ cm$^{-3}$) than knots C to E,
where \dent\ ranges from 1$\times$10$^3$ to 3$\times$10$^3$ cm$^{-3}$.

\subsubsection{Small-scale radial velocity along HH~223}

\begin{figure}
\centering
\resizebox{0.9\hsize}{!}{\includegraphics{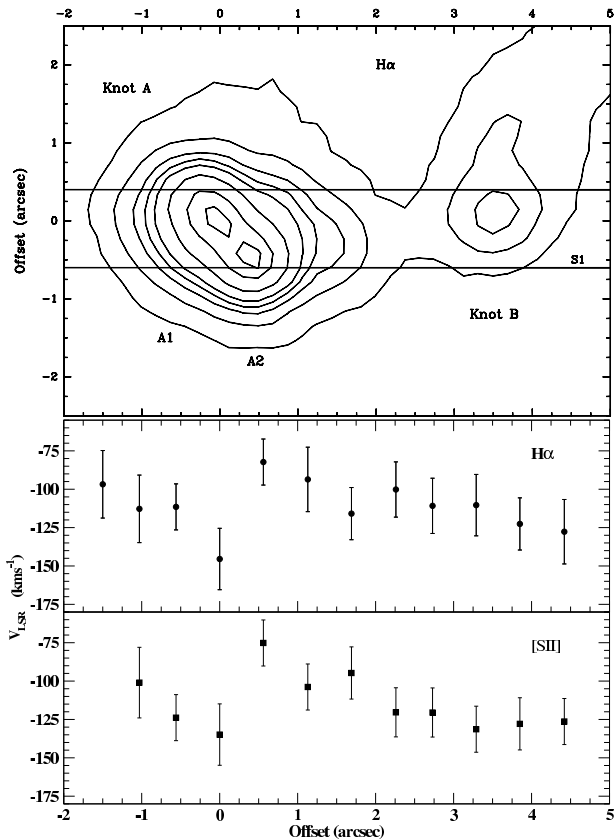}}
\caption{Lowel panel: Radial velocity along  knots A and B, derived from  
the H$\alpha$  and the \sii\ $\lambda$ 6716,
6731 \AA\ line centroids.
Upper panel: contours plot of the 
H$\alpha$ image and the S1 slit projection.}
\label{velab} 
\end{figure}

\begin{figure}
\centering
\resizebox{0.95\hsize}{!}{\includegraphics{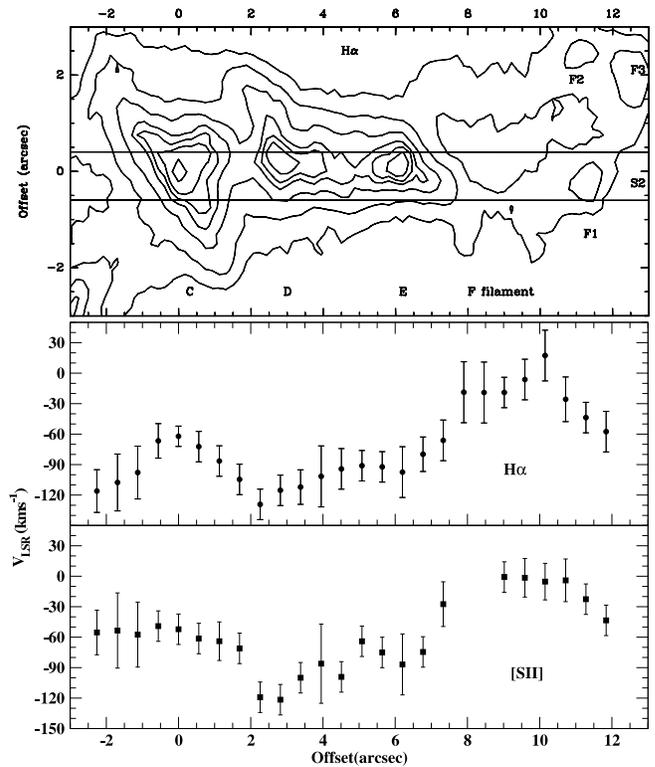}}
\caption{Same as Fig.\ \ref{velab} but for the slit position S2, 
intersecting
emission from knots C to F filament.}
\label{velcf} 
\end{figure}

The radial velocities along  HH~223, with a spatial sampling of $\sim$
0$\farcs$6, were derived from the line centroids of  Gaussian fits to the
H$\alpha$ and the \sii\ $\lambda$ 6716,6731 \AA\ lines. Results are displayed in Figs.\
\ref{velab} and \ref{velcf} for the slit positions S1 and S2 respectively.
From these figures,  the following trends are found: 
i) The velocity
behaviour as a fuction of the slit position is very similar for the lines 
considered. ii) The radial velocities derived appear highly blueshifted at all
the positions enclosed within the knots A to E and F1; 
velocities values slightly blueshifted  
or at the ambient gas velocity are found at positions corresponding to
the F filament, but out of the F1 faint knot. 
iii) In several regions, significant  variations in the
velocity can be found at scales  $\leq 1\arcsec$, smaller than 
the knot sizes.

Figure \ref{velab} displays the radial velocity behaviour along knots A and B.
There are appreciable velocity changes with position along knot A, the
velocities being more blueshifted towards the east. Such a behaviour is found
for all the emission lines (see also  Fig.\ \ref{plab}). We 
derived a difference in velocity of $\sim$ 60 \kms\ between two position within
knot A, which corresponds to 
the two emission peaks,
A1 and A2, of the narrow-band image 
(see also Fig.\ \ref{pvab}). The
radial velocity has only small variations along knot B, reaching  
similar values to those
found for positions eastwards of A1.

Figure \ref{velcf} displays the radial velocity behaviour along  S2. 
The more blueshifted velocities ($\sim-130$~\kms) are found 
for the knot D emission peak position,  reaching similar values to those
found for knot B.  Less blueshifted velocities
are found eastwards of knot D, reaching 
a value of $\sim-60$ \kms  around the knot C emission peak position. 
There is an
appreciable change in velocity for positions offset  $\sim +8\arcsec$ to
$+11\arcsec$ from
knot C, where the derived velocities are  compatible with the \vlsr\ value. 
Interestingly, the emission intersected by the slit at these positions mostly
comes from the  F filament, out of the three faint knots (F1-F3) found
engulfed in it. Thus we speculate that the emission in this region probably would
arise from ambient gas of a wall cavity which is already been excited and
dragged by the supersonic outflow.

Finally, it should be pointed out that, at several positions,  the line profiles
in all the strong emission lines of the unbinned spectra are suggestive of being
double-peaked. Due to the spectral resolution used,   we have not been able to
resolve with confidence the two velocity contributions suggested by these line
profiles. However, we believe that  the contribution from two velocity
components is reliable  and should be characterize from higher spectral
resloution data.

\section{Conclusions}

In  order to get knowledge on the kinematics and physical conditions in HH~223,
we performed long-slit spectroscopy at
two slit postions, covering  emission from all the bright knots (A to E) 
and from the low-brightness F filament.
 
The spectrum of each of the HH~223 bright knots was obtained by integrating the
emission within the slit window aperture encompasing the spatial extent of the 
knot  (ranging from
$\sim$ $2\arcsec$ to 3$\farcs$5).  The analysis of these spectra allowed us to 
characterize the nature of the knot emission. The  results are  summarized as
follows: 
\begin {itemize}  
\item  
For all the knots, the spectrum  appears as
produced by  shock-excited gas, characteristic of the HH objects.  
\item  
All the knot spectra have an intermediate/high degree of excitation, as 
their \sii\ and \nii\ to H$\alpha$ ratios indicate. 
 The estimated ionization
fraction (\fri) ranges from 0.15 in knot B to 0.3 in knot A.
The  electron density, derived from
the \sii\ 6716/6731 ratios, ranges from  \den\ $\simeq$ 240 (in knot C) to 2800
cm$^{-3}$.  The total density ranges from \dent $\ge$ 10$^4$ cm$^{-3}$ in
knots A and B, to \dent\ $\simeq$ 10$^3$ cm$^{-3}$ in knot E.
 Knot A is the densest and has the  highest excitation and ionization.  The lowest
excitation and ionization is found in knot B. 
\item  
The kinematics derived from
the emission lines indicates that the knot  emission is supersonic, with 
blueshifted velocities ranging from $-60$ to $-130$ ~\kms, the more
blueshifted velocity  values being found in knot B. Velocity values compatible 
with the ambient gas
are only derived for the F filament.  
\end{itemize} 

From the spectra obtained by binning the signal of each three adjacent pixels ($\sim$
0$\farcs$6)  along the slit, we searched for variations, both in the kinematics
and in the  physical conditions, at a scale smaller than the knot sizes. The
relevant results  are the following:
\begin {itemize} 
\item  
Reliable variations at a scale smaller than the knot sizes are found in
both, kinematics and physical conditions.  
\item 
Emission from knot A can be
resolved into two  substructures (A1 and A2), with a
projected offset of $\sim 0\farcs5$  between their  H$\alpha$ emission peaks,  
and a difference in velocity of $\sim$ 60~\kms, the
eastern (A1)  substructure being the most blueshifted one. 
\item 
Spatial variations  of \den\ up to  one order of magnitude are  found. The
lower \den\ values are found at positions where the slit mostly intersects gas
from the low-brightness filamentary nebula (F). The highest \den\ is found 
coinciding 
with the western substructure (A2) of  knot A. Other  enhancements of
\den\ are found, in addition, at several positions inside  knots C, D and
E. 
\end{itemize} 

In summary, concerning the nature of the emission, the 
spectroscopy let us to conclude that the emission from the HH~223
knots arises from shock-excited, highly blueshifted gas, being thus
characteristic  of an optical outflow having an inhomogeneous, knotty structure
in which each knot could enclose smaller subcondensations, as suggested by the
small-scale changes found in the kinematics and physical conditions. 
 The knots emission should then trace internal working surfaces of shocks
that originate because the supersonic gas is being ejected at varying speeds or
with different ejection directions, and faster ejecta overtake slower ones. A
variable ejection velocity and a wiggling knot pattern is consistent with the
exciting source being a YSO binary system. The H$_2$ emission detected at the
HH~223 location should arise from gas collisionally excited by shocks, which is
not dissociated because its lower velocity in the shock frame. Furthermore
most of the
emission from the  HH~223 low-brightness  optical nebula 
should arise from the gas excited
and dragged by the  CO outflow and traces an accelerated layer in the cavity
walls, since velocities compatible with the ambient gas 
velocity
are only derived  at positions  on 
the  lower-brightness optical nebula (F filament),  out of the F1 faint knot.

In order to complete the knowledge of  HH~223, it would be useful to map the whole extent of the emission with 
higher spatial  and spectral resolution,  searching for the knot substructures
and  resolving the contributions to the emission coming from the supersonic
outflow and from the dragged, excited nebular gas. In addition, spectra through
knot A, covering a wavelength range  including the \fe\ 13F and 14F
multiplets, are needed in order to properly derive \den\ around the A2 
substructure
from the \fe\ 7155/8617 ratio, suitable to perform a \den\ diagnostic when 
\den\ is  higher than $\sim$ 10$^4$ cm$^{-3}$, where the \sii\ 6716/6731 ratio
is not density sensitive.

\begin{acknowledgements} 
We acknowledge the support astronomer Amanda D. Anlaug  her help with the data
acquisition and Chin-Fei Lee for providing us the FITS file of
the CO emission  in L723.
The work of C.C.-G., R.E., R.L. and A.R. was supported by the Spanish MEC
grant
AYA2005-08523-C03 and the MICINN  grant AYA2008-06189-C03 (co-funded with FEDER
funds).
C.C.-G. acknowledges support from a MEC (Spain) FPU fellowship and from Junta
de Andaluc\'{\i}a (Spain).
R.L. acknowledges support from the OPTICON Access Programme.
ALFOSC is owned by the Instituto de Astrof\'{\i}sica de Andaluc\'{\i}a (IAA) 
and operated at the
NOT under agreement between IAA and the NBIfAFG of the Astronomical Observatory of Copenhagen.
 We thank the referee, John Bally, for his useful comments.
\end{acknowledgements}


\begin{thebibliography}{}

\bibitem[1996]{ang96} Anglada, G., Rodr\'{\i}guez, L.F., \& Torrelles, J.M., 1996, 
ApJL, 473, 123.


\bibitem[1999]{bac99} Bacciotti, F. \& Eisl\"offel, J 1999, A\&A 342, 717.


\bibitem[1981]{can81} Cant\'o, J., 1981 in Investigating the Universe, ed. F. Kahn
(Dodrecht:Reidel), 95.



\bibitem[2008]{car08} Carrasco-Gonz\'alez, C., Anglada, G., 
Rodr\'{\i}guez, L.F. et al., 2008, ApJ, 676, 1073.

\bibitem[2003]{est03} Estalella, R., Palau, A., Girart, J. M. et al., 
2003, Rev.\ Mex.\ Astron.\ Astrof.\ C. S., 15, 135.

\bibitem[2007]{gal07} G{\aa}lfalk, M. \& Olofsson, G., 2007, A\&A, 475, 281.

\bibitem[1997]{gir97} Girart, J. M., Estalella, R., Anglada, G. et al., 
1997, ApJ, 489, 743.


\bibitem[2008]{gir08} Girart, J. M., Rao, R. \& Estalella, R., 2008, ApJ, 
accepted.

\bibitem[1984]{gol84} Goldsmith, P.F., Snell, R.L., Hemeon-Heyer, M., \& Langer,
W.D., 1984, ApJ, 286, 599.
 
\bibitem[1994]{har94} Hartigan, P., Morse, J.A. \& Raymond, J., 1994, ApJ, 436,
125.

\bibitem[2002]{lee02} Lee, C.F., Mundy, L.G., Stone, J.M., \& Ostriker, E.C.,
2002, ApJ, 576, 294.

\bibitem[2006]{lop06} L\'opez, R, Estalella, R., G\'omez, G., Riera, A., 2006,
A\&A, 454, 233.

\bibitem[2009]{lop09} L\'opez, R., Acosta-Pulido, J.A. et al., 2009,
in preparation.

\bibitem[1982]{mea82} Meaburn, J., \& White, N.J., 1982, MNRAS, 199, 121.

\bibitem[1999]{pal99} Palacios, J., \& Eiroa, C., 1999, 346, 233.


\bibitem[1996]{rag96} Raga, A.C., B\"ohm, K.-H. \& Cant\'o, J., 1996, 
RevMexAA, 32, 161.

\bibitem[2003]{rie03} Riera, A., L\'opez, R., Raga, A.C., Estalella, R., 
Anglada, G., 2003, A\&A, 400, 213. 

\bibitem[2002]{shi02} Shirley, Y., Evans, N. J., \& Rawlings, J., 2002, 
ApJ, 575, 337.

\bibitem[1986]{tor86} Torrelles, J.M., Ho, P.T.P, Moran, J.M.,
 Rodr\'{\i}guez, L.F., Cant\'o J., 1986, ApJ, 307, 787.

\bibitem[1986]{vrb86} Vrba, F.J., Luginbuhl, C.B., Strom, S.E., Strom, K.M.\& Heyer, M.H.,
1986, AJ, 92, 633.

 


\end{thebibliography}
\end{document}